\documentclass[a4paper,12pt]{article}
\author 
{W.~X.~Ma
\thanks {On leave of absence from Institute of 
Mathematics, Fudan University, Shanghai 200433, China}$\ $
  and B.~Fuchssteiner\\
FB17, Mathematik-Informatik, Universit\"at-GH Paderborn,
\\ D-33098 Paderborn, Germany }
\title
{The Bi-Hamiltonian Structure of the Perturbation  Equations of KdV Hierarchy }
\date{\nonumber}  
\begin{document}
\maketitle
\begin{abstract}
The bi-Hamiltonian structure is established for the perturbation 
equations of KdV hierarchy and thus the perturbation 
equations themselves provide also examples among typical 
soliton equations. Besides, a more general bi-Hamiltonian 
integrable hierarchy is proposed and a
 remark is given for a generalization of the 
resulting perturbation equations to $1+2$ dimensions.
\end{abstract}




\def \la {\lambda}
\def \1 {\eta}
\def \2 {\varepsilon}
\def \La {\Lambda}
\def \be {\beta}
\def \al{\alpha}
\def \del{\delta}
\def \Del{\Delta}
\def \al{\alpha}
\def \vare{\varepsilon}

\def \part {\partial}

Perturbation theory is the study of the effects of small disturbances.
Its basic idea is to find approximate solutions to a concrete problem
by exploiting the presence of a small dimesionless parameter.
There have been a lot of works to investigate the perturbated soliton 
equations (see for example 
\cite{Herman} \cite{Matsuno} and references therein).
Tamizhmani and Lakshmanan have considered a perturbation effect of 
the unperturbated KdV equation 
and they have given rise to infinitely many Lie-B\"acklund symmetries and 
a Hamiltonian structure for the resulting equations
in Ref. \cite{TamizhmaniLakshmanan}. However they haven't obtained
 the corresponding bi-Hamiltonian formulation.

In this letter, we would like to discuss
the perturbation equations of the unperturbated whole KdV hierarchy, i.e.
the effect 
of the disturbance around solutions 
of  the following original KdV hierarchy 
\begin{equation}
u_{t_n}=\Phi^n (u)u_x,\ \Phi(u)=\part ^2 +2u_{x}\part ^{-1}+4u,\ \part
=\frac {d}{dx},\  n\ge 1.
\label{kdvh}
\end{equation}
The first equation is exactly the usual KdV equation
\begin{equation} u_{t_1}=u_{xxx}+6uu_x,\label{kdv}\end{equation}
which describes the unidirectional propagation of
long waves of small amplitude and has a broad of applications in a number of 
physical contexts such as hydromagnetic waves, stratified internal waves,
ion-acoustic waves, plasma physics and lattice dynamics (for details see, 
for example, \cite{Miura} \cite{AblowitzS}). 
The operator $\Phi (u)$ is a common hereditary recursion operator for 
the whole KdV hierarchy. 
We shall show that all of the resulting 
perturbation
equations possess a bi-Hamiltonian structure and thus they
constitute a typical integrable soliton hierarchy. We shall 
also point out a mistake on 
the Hamiltonian formulation in Ref. \cite{TamizhmaniLakshmanan}. 
Finally, a more general bi-Hamiltonian 
integrable soliton hierarchy is established and some
further discussion is presented.

We first consider the case of KdV equation 
(\ref{kdv}) and then consider the case of higher order KdV equations.
Let us make a perturbation expansion 
\begin{equation}
\hat{  u}=\sum_{i=0}^N \eta _i \2 ^i, \ \1 _i=\1 _i(x,t_1,t_2,\cdots)
,\ N\ge 1,\label{pexp}
\end{equation}
and call the equation 
\begin{equation}
\hat {  u}_{t_1} =\Phi (\hat {  u})\hat {  u}_x+\textrm{o}( \2 ^ N )
=\hat {  u}_{xxx}+6\hat {  u} \hat {  u}_{x}
 +\textrm{o}( \2 ^ N ),
\label{pkdv}
\end{equation}
the $N$-th order perturbation equation of KdV equation (\ref{kdv}).
It is easy to find that 
\[
\Phi (\hat{u})=\sum_{i=0}^N\Phi (\1 _i)\2 ^i,\ 
\Phi (\1 _i)=\del _{i0}\part ^2 + 2\1 _{ix}\part ^{-1}+4 \1 _{i},\ 
 0\le i\le N,
\]
where $\delta _{i0}$ is the Kronecker's symbol.
Therefore the $N$-th order perturbation equation (\ref{pkdv}) may be 
rewritten as  
\[
\hat {  u}_{t_1}
\equiv \Bigl ( \sum_{i=0}^N
\Phi (\1 _i)\2 ^i\Bigr )\Bigl ( \sum_{i=0}^N \1 _{ix}\2 ^i\Bigr )\ \ (
\textrm{mod}\,\2 ^{N+1}).\]
A balance of the coefficients of like powers of $\2 $
 leads to the following equivalent equation
\begin{equation}
\eta _{t_1}= \hat {  \Phi} ( \eta )\eta _x,\ \1 =(\1 _0,\1 _1,\cdots,\1 _N)^T,
 \label{cpkdv}\end{equation} 
where the operator $\hat {  \Phi} (\1 )$ reads as
\begin{equation}\hat {  \Phi}=\hat {  \Phi} (\eta )=
\left [\begin{array}{cccc}
\Phi (\1 _0) & & & 0\\
\Phi (\1 _1) & \ddots & &  \\
\vdots & \ddots & \ddots &  \\
\Phi (\1 _N) & \cdots &\Phi (\1 _1) & \Phi (\1 _0)
\end{array} \right ].\label{ho}
\end{equation}
By its first component of (\ref{cpkdv}), we see 
that the perturbation expansion
(\ref{pexp}) is an expansion around an exact solution $\eta _0$
 of KdV equation (\ref{kdv}).
Moreover $\hat {u}$ is an approximate solution of KdV equation
(\ref{kdv}) to a precision 
$\textrm{o}(\2 ^{N})$ when $\eta $ satisfies (\ref{cpkdv}).
However our aim here is not to get some approximate solutions.
What we want to discuss is some algebraic structures that the perturbation 
equations possess.

Let us introduce two differential operators
\begin{equation}
\hat {  M}=\left [ \begin{array}{cccc}
0 & &   & J(\1 _0)\\
 &  &\vdots   & J(\1 _1) \\
 &\vdots   & &\vdots   \\
J(\1 _0) & J(\1 _1) &\cdots &J(\1 _N) 
\end{array} \right ],\ \hat {  J}=\left [ \begin{array}{cccc}
0 & & & \part \\
 &  &\part &  \\
 & \vdots &  &  \\
\part  &  & & 0 
\end{array} \right ]
\end{equation}
with $J(\1 _i)=\delta _{i0}\part ^3 +2\1 _{ix} +4\1 _i\part,\ 0\le i\le N$.
Note that the operator 
$\hat {  M}+ \al \hat {  J}$
is a Hamiltonian operator for any constant 
$\al $ (see for example Ref. \cite{Mawx})
 and that 
\[\hat {  \Phi} (\1 )= \hat {  M} \hat {  J }^{-1},\ \hat {  J }
\hat {  \Psi}=\hat {  \Phi}\hat {  J },\ \hat {  \Psi }=\hat {  \Phi}^*,\]
where the asterisk appended to $\hat {  \Phi}$
 denotes the conjugate operation.
 Therefore $\hat {J},\hat {M}$ constitute a Hamiltonian pair and further
the operator $\hat {  \Phi} (\1 )$ is 
hereditary \cite{Fuchssteiner},\cite{FuchssteinerFokas}.

We are now in the position to 
construct a bi-Hamiltonian structure \cite{Magri} \cite{GelfandDorfman}
for the perturbation equation
(\ref{cpkdv}). 
 We first have the first Hamiltonian structure
\begin{equation}
 \1 _{t_1}=\hat {  \Phi} (\1 ) \1 _x= \hat {  \Phi} (\1 ) \hat {  J}
\frac {\delta H_0}{\delta  \1 }=\hat {  M}
\frac {\delta H_0}{\delta  \1 },\ H_0=H_0(\1 )=
\frac12 \sum_{i=0}^N\1 _i\1 _{N-i}.
\end{equation}
Second noting that 
\[\Psi (\1 _i):=\Phi ^*(\1 _i)=\delta _{i0}\part ^2+2\1 _i +2\part ^{-1}
\1 _i\part,\ 0\le i\le N,\]
 we can compute
\begin{eqnarray} f_1(\1 )&:=&
\hat {  \Psi} (\1 )
\frac {\delta H_0}{\delta  \1 }=\left[\begin{array}{cccc}
\Psi(\1 _0)
 &\Psi(\1 _1)&\cdots&\Psi(\1 _N)\\
 & \ddots &\ddots &  \vdots\\
 & & \ddots&\Psi(\1 _1)
\\
0  &  & & \Psi(\1 _0)
\end{array} \right ]
\left[\begin{array}{c}
\1 _N\\ \1 _{N-1}\\ \vdots \\ \1 _1 \\ \1 _0
\end{array}\right ]\vspace{2mm}\nonumber
\\ 
&=&\left[\begin{array}{c}
\1 _{Nxx}+3\sum_{\textrm {
{\tiny $
\begin{array}{c} i+j=N\\ i,j\ge 0\end{array}$}}}\1 _i\1 _j\\
\1 _{N-1,xx}+3\sum_{\textrm {
{\tiny $
\begin{array}{c} i+j=N-1\\ i,j\ge 0\end{array}$}}}\1 _i\1 _j\\
\vdots \\
\1 _{0xx}+3\1 _0^2
\end{array}\right ]=\frac {\delta H_1}{\delta  \1 },\nonumber
\end{eqnarray}
where the Hamiltonian function $H_1$ is defined by
\[
H_1=\int _0^1<f_1(\lambda \1 ), \1 >d\lambda=\frac 12 
\sum_{\textrm {
{\scriptsize $
\begin{array}{c} i+j=N\\ i,j\ge 0\end{array}$}}}\1 _i\1 _{jxx}+
\sum_{\textrm {
{\scriptsize $
\begin{array}{c} i+j+k=N\\ i,j,k\ge 0\end{array}$}}}\1 _i\1 _{j}\1 _k.\]
Here and hereafter $<\cdot,\cdot>$ stands for the standard inner on $R^{N+1}$.
The above analysis allows us to conclude
that  the perturbation equation (\ref{cpkdv}) possesses a bi-Hamiltonian 
structure
\begin{equation}
\eta _{t_1}= \hat {  \Phi} ( \eta )\eta _x=\hat {  J}\frac
 {\delta H_{1}}{\delta  \1 }=\hat {  M}\frac {\delta H_{0}}{\delta  \1 }.
\end{equation}
When $N=1,\ N=2$, the perturbation equation (\ref{cpkdv}) becomes
\[\left[
\begin{array}{c}
\eta _0\\ \eta _1
eta _0\eta _{0x}\\
\eta _{1xxx}+6(\eta _0\eta _{1})_x
\end{array}\right ],\]
\[\left[
\begin{array}{c}
\eta _0\\ \eta _1\\ \eta _2
\end{array}\right ]_{t_1}=\left[
\begin{array}{c}
\eta _{0xxx}+6\eta _0\eta _{0x}\\
\eta _{1xxx}+6(\eta _0\eta _{1})_x\\
\eta _{2xxx}+6\eta _1\eta _{1x}+6(\eta _0\eta _{2})_x
\end{array}\right ],\]
respectively. These two equations are all
integrable coupling ones with KdV equation. 
Integrable coupling is an interesting subject 
in soliton theory
\cite{Fuchssteiner2}. Note that
the second equation may be reduced to the first one if we take 
$\eta _1=0$. 

For the case of higher order KdV  equations,
by induction we  can  engender the following equivalent equation
\begin{equation}
\eta _{t_n}=\hat {  \Phi} ( \eta )\eta _{t_{n-1}}= \hat 
{  \Phi} ^n( \eta )\eta _x \ (n>1)
\label{cpkdvn}
\end{equation}
to the $N$-th order perturbation equation of  the  $n$-th KdV equation
 \begin{equation}
\hat {  u}_{t_n} =\Phi ^n(\hat{u}) \hat{u}_x+\textrm{o}( \2 ^ N ),
\label{pkdvn}
\end{equation}
namely, 
\[
\hat {  u}_{t_n} \equiv 
\Phi ^n(\hat{u}) \hat{u}_x\equiv \Bigr (\sum_{i=0}^N\Phi (\eta _i)
\2 ^i\Bigl ) 
\Bigr (\sum_{i=0}^N \eta _{it_{n-1}}\2 ^i    \Bigl ) 
 \ \ (\textrm{mod}\, \2 ^ {N+1} ).\]
According to the bi-Hamiltonian theory (for a detailed description see
\cite{Olver}, \cite{Schmidt} 
and \cite{Casati}),
we know the vectors
$\hat {  \Psi }^n(\1 )\frac {\delta H_0}{\delta  \1 }
$, $n\ge 1$,
are all gradient vectors because 
$\hat {  \Psi }(\1 )\frac {\delta H_0}{\delta  \1 }$ is a gradient vector
and $\hat {\Psi }^*=\hat {\Phi }$ is hereditary. Therefore
 there exist functions $H_n$
so that 
\[f_n(\1 ):=
\hat {  \Psi }^n(\1 )\frac {\delta H_0}{\delta  \1 }
=\frac {\delta H_n}{\delta  \1 },\ n\ge 1.\]
Moreover the Hamiltonian functions can be computed by
\[ H_n=H_n(\1 )=\int_0^1<f_n(\lambda \1 ), \1 >d\lambda.\]
Therefore we see that the perturbation equation (\ref{cpkdvn})
of the $n$-th KdV equation in the hierarchy (\ref{kdvh})
possesses a bi-Hamiltonian structure
\begin{equation}
\eta _{t_n}= \hat {  \Phi} ^n( \eta )\eta _x=\hat {  J}\frac
 {\delta H_{n}}{\delta  \1 }=\hat {  M}\frac {\delta H_{n-1}}{\delta  \1 }.
\label{biham}\end{equation}
In fact, the perturbation equation (\ref{cpkdvn}) has a multi-Hamiltonian
structure 
\[
\eta _{t_n}=\hat {  J}\frac
 {\delta H_{n}}{\delta  \1 }=\hat {  J}\hat {  \Psi}
\frac {\delta H_{n-1}}{\delta  \1 }=\cdots =
\hat {  J}\hat {  \Psi}^n
\frac {\delta H_{0}}{\delta  \1 },\]
where $\hat{J}\hat{\Psi}^i,\,0\le i\le n$, are all Hamiltonian operators and 
constitute Hamiltonian pairs with each other. It is also interesting 
to note the hereditary operator $\hat {\Phi }^n(\1 )$ has the following 
concrete form
\[\hat {\Phi }^n(\1 )=\left[\begin{array}{cccc}
A _0& & &0\\
A_1&\ddots& & \\
\vdots & \ddots &\ddots & 
\\ A_N &\cdots & A_1 &A_0
\end{array}\right],\]
where the operator $ A_j,\ 0\le j\le N,$ are determined by
\[A_j=\sum_{\begin{array}{c}
i_1+\cdots +i_n=j\\ i_1,\cdots i_n\ge 0\end{array}}
\Phi(\1 _{i_1})\cdots \Phi(\1 _{i_n}),\ 0\le j\le N.\]
Unlike the hereditary operator for the Kepler system \cite{MarmoVilasi},
the above hereditary operator can generate new independent 
constants of motion $H_n$
from a starting one $H_0$.

By now, we have proposed a bi-Hamiltonian structure for the
perturbation equations of the whole KdV hierarchy. This bi-Hamiltonian
structure yields infinitely many symmetries and  infinitely many
corresponding 
constants of motion in involution 
for every perturbation equation and thus
the perturbation equations of KdV hierarchy
 provide a typical hierarchy of soliton equations.
 It is worth noting that the algebraic 
structures of the  first order perturbation equations may also
 be well described by the  
perturbation bundle \cite{Fuchssteiner2}
and cotangent bundle \cite{Walter}.

We  remark that some cases, especially $N=1$, for KdV equation
 (\ref{kdv})
has in detail been considered in Ref.\cite{TamizhmaniLakshmanan}.
But they only described a Hamiltonian structure. Moreover that Hamiltonian 
structure is not suitable for the perturbation equations of  KdV equation,
which leads to some confusion of the symbols. 
For example, in the case of $N=1$, we cannot write
\[
\1 _{t_1}=\hat{\Phi }\1 _x=
\left[ \begin{array}{cc}
\part & 0\\0&\part 
\end{array}\right]\frac {\delta H_0'}{\delta \1 }.
\]
Actually, there doesn't exist this kind of Hamiltonian function $H_0'$,
because
$(\1 _{0xx}+3\1 ^2,\1 _{1xx}+6\1 _0\1 _1)^T$ is not a gradient vector.
However this is a small problem, which is easily changed to the correct 
version mentioned above. 

There also exists a more general result than the bi-Hamiltonian structure 
(\ref{biham}). Let 
$a_s ,\ 1\le s \le p,$ $b_s ,\ c_s ,\ 1\le s \le q,$
$d_s ,\ 1\le s \le r,$ are arbitrary real constants, and 
$k_s ,\ 1\le s \le p,$ $l_s\ 1\le s \le q,$ 
$m_s ,\ 1\le s \le r,$ are distinct non-negative integers not greater than $N$.
We further choose that $\tilde J = \hat J$ and 
\begin{equation}
\tilde {  M}=\left [ \begin{array}{cccc}
0 & &   &\tilde J(\1 _0)\\
 &  &\vdots   &\tilde J(\1 _1) \\
 &\vdots   & &\vdots   \\
\tilde J(\1 _0) & \tilde J(\1 _1) &\cdots &\tilde J(\1 _N) 
\end{array} \right ],
\end{equation}
where 
the operator $\tilde J (\eta _i),\ 0\le i\le N$, are defined by
\begin{eqnarray} \tilde J (\eta _i)& =&
 \sum_{s =1}^p a_s \delta _{ik_s }  \partial ^3 
+ \sum_{s =1}^q( b_s \delta _{il_s }\partial ^3 
+ c_s \delta _{il_s }\partial )
+
\sum_{s =1}^r d_s \delta _{im_s }  \partial \nonumber \\ &&
+2\eta _{ix}+4\eta _i\partial,\ \ 0\le i\le N.\label{tJ}
\end{eqnarray}
Note that (\ref{tJ}) means we choose 
\[ \begin{array}{l}\tilde J (\eta _{k_s }) =a_s 
\partial ^3 +2\eta _{k_s ,x}+4\eta _{k_s }\partial ,\ 1\le s \le p,\vspace{2mm}\\
\tilde J (\eta _{l_s }) =b_s 
\partial ^3 +c_s 
\partial +2\eta _{l_s ,x}+4\eta _{l_s }\partial ,\ 1\le s \le q,\vspace{2mm}\\
\tilde J (\eta _{m_s }) =d_s 
\partial  +2\eta _{m_s ,x}+4\eta _{m_s }\partial ,\ 1\le s \le r,\vspace{2mm}\\
\tilde J (\eta _{i})=2\eta _{ix}+4\eta _{i}\partial ,\ i\,
/ \hspace{-0.9em} \in 
 \{k_s ,\ 1\le s \le p;\ l_s ,\ 1\le s \le q;\ m_s ,\ 1\le s \le m\}.
\end{array} \]
The differential operators $ \tilde J$ and $\tilde M$ are still a Hamiltonian 
pair \cite{Mawx} and hence 
the operator $\tilde  {  \Phi} (\1 )= \tilde  {  M}\tilde  {  J }^{-1}$ is 
a hereditary symmetry operator. 
This leads to the following hierarchy of integrable equations which 
possesses a more general 
bi-Hamiltonian structure
 \begin{equation}\eta _{t_n}=\tilde {\Phi} ^n\eta _x
 =\tilde J \frac {\delta \tilde H_n}{\delta  \1 }=
\tilde M \frac {\delta \tilde H_{n-1}}{\delta  \1 },\ n\ge 1\label{morebiham}
\end{equation}
with  the Hamiltonian functions defined by
\[ \tilde H_n=\tilde H_n(\1 )=\int_0^1<(\tilde {\Phi} ^n\eta _x)
 (\lambda \1 ), \1 >d\lambda ,\ n\ge 0, \]
in which $\tilde H_0$ and $\tilde H_1 $ 
read as
\begin{eqnarray} &&\tilde H_0=H_0=
\frac12 \sum_{i=0}^N\eta _i\eta _{N-i},\nonumber\\ &&
\tilde H_1=
\frac12 \sum_{s=1}^pa_s
\sum_
{\textrm {
{\scriptsize $\begin{array}{c}
i+j=N-k_s\\ i,j\ge 0\end{array}$}}}
\eta _i\eta _{jxx} \nonumber\\ &&+
\frac12 \sum_{s=1}^q\Bigr(b_s
\sum_{\textrm {
{\scriptsize $\begin{array}{c}
i+j=N-l_s\\ i,j\ge 0\end{array}$}}}
\eta _i\eta _{jxx}+c_s
\sum_{\textrm {
{\scriptsize $\begin{array}{c}
i+j=N-l_s\\ i,j\ge 0\end{array}$}}}
\eta _i\eta _{j}\Bigl)\nonumber\\ &&
+
\frac12 \sum_{s=1}^rd_s
\sum_{\textrm {
{\scriptsize $\begin{array}{c}
i+j=N-m_s\\ i,j\ge 0\end{array}$}}}
\eta _i\eta _{j}+
\sum_{\textrm {
{\scriptsize $
\begin{array}{c} i+j+k=N\\ i,j,k\ge 0\end{array}$}}}
\1 _i\1 _{j}\1 _k.\nonumber
\end{eqnarray}
Evidently, (\ref{biham}) is a special case of the hierarchy 
(\ref{morebiham}), one reduction with $k_1=0$ and $a_1=1,\ a_s=0,\ s\ne 1$,
$b_s=c_s=d_s =0,\ \forall s $.

We point out that our method can also be applied to discussing 
the forced integrable 
systems \cite{Kaup}. It is sufficient to further make the perturbation of 
value boundary conditions and then we can get new forced integrable 
systems. The resulting forced systems can be solved 
by the same approach as the one in the original 
forced integrable systems \cite{Kaup}. However in the present letter, we
haven't considered this kind of interesting problems due to the limited 
space.

Finally, we suggest that it may be more fruitful to think of the 
perturbation expansion $\hat {u}$ as depending on two independent 
space variables, $x$ and $y=\2 x$, i.e.
\begin{equation}
 \hat{u}=\sum_{i=0}^N\eta _i(x,y,t_1,t_2,\cdots)\2 ^i,\ y=\2 x.\label{twov}
\end{equation}
This kind of expansions comes from the perturbation method of multiple
scales \cite{Nayfeh} and the variable $y$ is called 
a slow variable. The perturbation method of multiple
scales for time variable has been applied to the study of perturbated
soliton equations (see for instance \cite{Matsuno}) and the study of 
solutions to soliton equations (see for instance \cite{last}).
Under the expansion (\ref{twov}), we have
\[ \frac {d}{dx}=\frac {\part }{\part x}+\2 \frac {\part }{\part y}.\]
Therefore from the $N$-th order perturbation equation
of KdV equation (\ref{kdv}), we can similarly obtain the following equation
in $1+2$ dimensions
\[\left \{
\begin{array}{l}
\1 _{0t_1}=\1 _{0xxx}+6\1 _0\1 _{0x},\\
\1 _{1t_1}=\1 _{1xxx}+3\1 _{0xxy}+6(\1 _0\1 _{1})_x+6\1 _0\1 _{0y},\\
\1 _{2t_1}=\1 _{2xxx}+3\1 _{1xxy}+3\1 _{0xyy}+6(\1 _0\1 _{2})_x
+6\1 _1\1 _{1x}+6(\1 _0\1 _{1})_y,\\
\1 _{jt_1}=\1 _{jxxx}+3\1 _{j-1,xxy}+3\1 _{j-2,xyy}+\1 _{j-3,yyy}
\\ \ \qquad +6\Bigl(\sum_{i=0}^j\1 _i\1 _{j-i,x}+\sum_{i=0}^{j-1}
\1 _i\1 _{j-i-1,y}
\Bigr),\ 3\le j\le N.\end{array}\right.\]
This is a generalization of the perturbation equations 
of KdV hierarchy.
However, we don't know the complete answer on the integrability of the above
equation, even in the case $N=1$ where 
the above equation reads as 
\[\left \{\begin{array}{l}
\1 _{0t_1}=\1 _{0xxx}+6\1 _0\1 _{0x},\\
\1 _{1t_1}=\1 _{1xxx}+3\1 _{0xxy}+6(\1 _0\1 _{1})_x+6\1 _0\1 _{0y}.
\end{array}\right.\]
This equation is also a coupling one with KdV equation.
If integrable, it provides us an another example 
which gives an integrable coupling equation
with KdV equation. 

\vskip 5mm

\noindent{\bf Acknowledgments:} 
The authors acknowledge valuable comments of the referee and 
helpful discussions with W. Oevel. One of the authors (W. X. Ma) 
would like to thank the 
Alexander von Humboldt Foundation, 
the National Natural 
Science Foundation of China and Shanghai Science and Technology Commission
of China
 for financial support.

\end{document}